\documentclass{svjour3}                     
\smartqed  
\usepackage{graphicx}
\usepackage{hyperref}
\usepackage{mathptmx}      
\newcommand\al{\alpha}

\newcommand\ga{\gamma}

\newcommand\ka{\kappa}
\newcommand\la{\lambda}

\newcommand\ta{\tau}

\newcommand\ps{\psi}
\newcommand\om{\omega}

\newcommand\mn{{\mu\nu}}

\newcommand\half{{\textstyle{1\over 2}}}
\newcommand\quar{{\textstyle{1\over 4}}}

\newcommand\etal{{\it et al.}}

\newcommand\pt[1]{\phantom{#1}}
\newcommand\ol[1]{\overline{#1}}

\newcommand\nsc[3]{\om_{#1}^{{\pt{#1}}#2#3}}

\newcommand\ivb[2]{e^{#1}_{{\pt{#1}}#2}}
\newcommand\uvb[2]{e^{#1#2}}

\newcommand\ab{\overline{a}}

\def\twiddle{\lower4pt\hbox{\hskip-0pt{$\widetilde{}$}}}
\def\m@th{\mathsurround=0pt}
\def\cmapstochar{\mathrel{\rlap{
  \lower0.1pt\hbox{\hskip-1.75pt{$\mapstochar$}}}
  \raise0pt\hbox{\hskip2.5pt{$\twiddle$}}}}
\def\notsimfill{$\m@th\cmapstochar$}
\def\scroodle#1{\vbox{\ialign{##\crcr\notsimfill\crcr
  \noalign{\kern-4pt\nointerlineskip}
   $\hfil\displaystyle{#1}\hfil$\crcr}}}

\def\cmapstocharbig{\mathrel{\rlap{
  \lower0.1pt\hbox{\hskip0.25pt{$\mapstochar$}}}
  \raise0pt\hbox{\hskip4.5pt{$\twiddle$}}}}
\def\notsimfillbig{$\m@th\cmapstocharbig$}
\def\scroodlebig#1{\vbox{\ialign{##\crcr\notsimfillbig\crcr
  \noalign{\kern-4pt\nointerlineskip}
   $\hfil\displaystyle{#1}\hfil$\crcr}}}

\newcommand\X{t_{\la \mn \ldots}}
\newcommand\Xb{\overline{t}_{\la \mn \ldots}}
\def\Xtw{\scroodle{t}_{\la \mn \ldots}}

\newcommand\atw{\scroodle{a}}

\newcommand\lrpartial{\raise 1pt\hbox{$\stackrel\leftrightarrow\partial$}}

\newcommand\lrDmu{\stackrel{\leftrightarrow}{D_\mu}}

\newcommand{\beq}{\begin{equation}}
\newcommand{\eeq}{\end{equation}}
\newcommand{\bea}{\begin{eqnarray}}
\newcommand{\eea}{\end{eqnarray}}
\newcommand{\bit}{\begin{itemize}}
\newcommand{\eit}{\end{itemize}}

\journalname{Hyperfine Interactions}
\begin{document}

\title{Gravitational physics with antimatter
}

\author{Jay D.\ Tasson}

\institute{Jay D.\ Tasson \at
              Physics Department\\
	      Indiana University\\
	      Bloomington, IN 47405 U.S.A.\\
              \email{jtasson@indiana.edu} 
}

\date{Published: 28 August, 2009.  The original publication is available
at \href{http://www.springerlink.com/content/e3533787v4577h3l/}{www.springerlink.com}.}

\maketitle

\begin{abstract}
The production of low-energy antimatter 
provides unique opportunities 
to search for new physics in an unexplored regime. 
Testing gravitational interactions with antimatter
is one such opportunity. 
Here a scenario based on Lorentz and CPT violation
in the Standard-Model Extension
is considered
in which anomalous gravitational effects in antimatter
could arise.
\end{abstract}

\section{Introduction}
\label{intro}

Antimatter is a regime
in which numerous predictions
of our current best description of nature,
the Standard Model of particle physics
together with Einstein's General Relativity,
remain untested.
As such, 
experiments with antimatter provide an opportunity
to detect new physics,
and are required to place our existing theories
on an experimental foundation
in this regime.

One such property
that deserves experimental testing
is the interaction of antimatter with gravity.
Within our present theories,
CPT symmetry,
which has passed all tests to date \cite{tables}
indicates that the properties of an antiparticle system
should be the same as those of a particle system.
Thus,
CPT symmetry suggests that a pair of antiparticles
should interact gravitationally
in the same manner as a pair of particles.
However,
arguing from CPT symmetry alone
provides no information about the gravitational interaction
of an antiparticle with a particle.
Moreover,
there exist ways in which CPT could be violated \cite{cpt07,sme,akgrav,mav},
some of which have never been tested.
Of these untested potential violations,
some could be large
relative to existing bounds \cite{akjtshort}.

Numerous attempts have been made 
to place indirect limits 
on the degree to which the behavior of an antimatter test particle
in a gravitational field
may differ from that of a matter particle \cite{nieto}.
However,
such considerations
are inherently model dependent,
and it is still possible
to find models
that predict an appreciable difference
in the gravitational response of matter and antimatter.
One such suggestion
is a Supergravity inspired model
involving vector and scalar partners to the graviton \cite{scherk}.
In the remainder of these proceedings,
the possibility that such differences
can be produced by violations of Lorentz and CPT symmetry
in the context of the Standard-Model Extension (SME) \cite{sme,akgrav}
is outlined.
This work was done in collaboration 
with Alan Kosteleck\'y \cite{akjtshort,lvgap}.

\section{Standard-Model Extension}

The SME is a general theoretical framework
for testing Lorentz and CPT symmetry.
It contains the Standard Model and General Relativity
along with arbitrary coordinate-independent Lorentz violation
in the form of an effective quantum field theory \cite{sme,akgrav}.
Searches for Lorentz violation
within the SME
are motivated in part
by the fact that although General Relativity
and the Standard Model provide an excellent description of nature
at our present low energies,
a single quantum-consistent theory
at the Planck scale is lacking.
Ideally experiment would guide the development
of the underlying theory;
however,
direct experiments at the Planck scale are infeasible.
Lorentz and CPT violation
offer the possibility of detecting suppressed Planck-scale effects
with existing technology \cite{ks}
and have been shown to be possible
within the context of numerous candidates
for the underlying theory.

A large number of experimental searches
for Lorentz violation
have been performed
in the context of the minimal SME.
In the Minkowski-spacetime limit
those tests include
experiments with
electrons \cite{eexpt,eexpt2,eexpt3,eexpt4,eexpt5,eexpt6},
protons and neutrons \cite{pnexpt,pnexpt2},
photons \cite{photonexpt,photonexpt2,photonexpt3},
mesons \cite{hadronexpt,hadronexpt2},
muons \cite{muexpt,muexpt2},
neutrinos \cite{nuexpt,nuexpt2},
and the Higgs \cite{higgs}.
Gravitational searches have also begun \cite{gravexpt}
based on investigations of the pure gravity sector \cite{lvpn}
and gravitational couplings in the matter sector \cite{akjtshort,lvgap}.
While the experimental searches performed thus far
have revealed no compelling
experimental evidence
for Lorentz violation,
much remains unexplored.
For example,
only about half
of the coefficients for Lorentz violation
in the minimal SME
involving light and ordinary matter
(protons, neutrons, and electrons)
have been investigated experimentally,
and other sectors
remain nearly unexplored.

\section{Relativistic Theory}

Here we consider the QED limit of the SME
with gravitational couplings \cite{akgrav}.
The general geometric framework assumed
is Riemann-Cartan spacetime, 
which contains the Riemann curvature tensor $R^\ka_{\pt{\ka} \la \mu \nu}$
as in General Relativity
and allows for a nonzero torsion tensor $T^\la_{\pt{\la} \mu \nu}$,
the effects of which
have been tightly constrained
by a recent re-interpretation
of bounds on SME coefficients
as bounds on torsion \cite{krt}.
The basic non-gravitational fields
are the photon $A_\mu$
and the Dirac fermion $\ps$,
while the spin connection $\nsc \mu a b$
and the vierbein $\ivb \mu a$
are taken as fundamental
gravitational objects.

The action of the minimal SME
can be expanded in the following way:
\beq
S = S_G + S_\ps + S^\prime.
\label{SMEaction}
\eeq 
The action of the pure-gravity sector
is provided by the first term $S_G$.
It contains the dynamics
of the gravitational field
and can also contain
coefficients for Lorentz violation
in that sector \cite{akgrav,lvpn}.
The second term provides
the action for the fermion sector,
which takes the form
\beq
S_\ps
= 
\int d^4 x \Big[\half i e \ivb \mu a \ol \ps 
\left( \ga^a - c_{\mu\nu} \uvb \nu a \ivb \mu b \ga^b - \ldots \right) \lrDmu \ps 
- e \ol \ps \left( m + a_\mu \ivb \mu a \ga^a + \ldots \right) \ps\Big]. 
\label{qedxps}
\eeq
The symbols
$a_\mu$ and $c_\mn$
are coefficient fields for Lorentz violation
of the minimal fermion sector
included here as a sample of such terms.
The ellipsis here
contains additional coefficients fields for Lorentz violation.
In general,
these fields vary with position
and
differ for each species of particle.
Note also that $a_\mu$
is CPT violating,
while $c_\mn$ is CPT preserving
and can be taken as traceless.
For additional discussion
of the fermion-sector action,
see Ref.\ \cite{akgrav}.

The final portion
of the action, 
$S^\prime$, 
contains the dynamics associated 
with the coefficient fields for Lorentz violation
and is responsible for spontaneous Lorentz-symmetry breaking.
Through spontaneous symmetry breaking,
the coefficient fields for Lorentz violation
are expected to
obtain vacuum values.
Thus,
it is possible
to express an arbitrary coefficient field for Lorentz violation
$\X$ as,
$\X = \Xb + \Xtw$,
where 
$\Xb$ is the corresponding vacuum value
and $\Xtw$
represents the fluctuations
about that vacuum value.
In general these fluctuations contain both
the massless Nambu-Goldstone modes \cite{lvng}
and massive modes \cite{lvmm}
associated with Lorentz-symmetry breaking.
The necessary tools to analyze fermion experiments
in the presence of gravity and Lorentz violation
may be developed
without specifying $S^\prime$ \cite{lvgap}.
For the $a_\mu$ case
those results are obtained in Ref.\ \cite{akjtshort}
and are summarized in the following section.

\section{Non-Relativistic Analysis}

One can proceed from the fully relativistic action (\ref{qedxps})
toward experimental analysis at a number of different levels \cite{lvgap}.
Relativistic quantum-mechanical investigations
can be performed by obtaining the relativistic quantum hamiltonian
from action (\ref{qedxps}).
Doing so involves a redefinition of the fermion field
followed by the usual Euler-Lagrange procedure.
The resulting hamiltonian
can be used to obtain the non-relativistic physics
via a Foldy-Wouthuysen transformation.
The classical lagranian
can then be obtained by inspection
of the non-relativistic hamiltonian,
which allows one to replace $S_\ps$ in Eq.\ (\ref{SMEaction})
with
\beq
S_u = \int d\ta \left(-m  \sqrt{-g_\mn u^\mu u^\nu}
- a_\mu u^\mu \right),
\eeq
for the purposes of analyzing classical point-particle experiments
at leading order in Lorentz violation
associated with $a_\mu$
and leading order in the metric fluctuation.

A generic investigation
of spontaneous breaking
performed by requiring coordinate independence of the physics
and geometric compatibility
provides the form of the fluctuations
in the Nambu-Goldstone limit.
Under such circumstances,
the fluctuations,
can be written in terms of the metric fluctuation and the vacuum value.
In the case of $a_\mu$ for example \cite{akjtshort},
\beq
\atw_\mu = \half \al \ab^\mu h_\mn - \quar \al \ab_\mu h^\nu_{\pt{\nu} \nu}
\eeq
in harmonic coordinates,
where $\al$ is determined by the strength of the coupling
to gravity.
Similarly,
by restricting to theories
for which Newton's third law holds,
it is possible to find the Lorentz-violating contributions to the metric fluctuation,
which
in the $a_\mu$ case take the form
$h_{00} \supset 4 \al \ab_0 U/m$,
in terms of the Newtonian potential $U$.

\section{Experiments}

Numerous suggestions
for experiments that can achieve sensitivity 
to $\ab_\mu$ are considered in Refs.\ \cite{akjtshort,lvgap},
and several experimental constraint are obtained
on combinations
of the 12 $\ab_\mu$ coefficients for ordinary matter.
The possibility of scenarios in which the CPT odd effects
of $\ab_\mu$ considered explicitly in that work
could cancel against CPT even Lorentz-violating effects
within the SME is also noted.
Under such circumstances
gravitational tests with antimatter
could disentangle the situation.

Several proposals have been made
to perform gravitational tests with neutral antimatter
or neutral combinations of particles and antiparticles.
These include tests
with  antihydrogen,
antineutrons,
muonium,
and positronium.

Tests with antineutrons
are mentioned in the proposal of Ref.\ \cite{antiinterf}.
Such tests
are of potential interest here since they
could provide a clean bound on neutron coefficients.
However they are technically challenging at present
due to issues associated with controlling
the particles.

The proposal of Ref.\ \cite{positronium}
suggests that interferometric methods
could be used to obtain sensitivity
to the gravitational acceleration
of a positronium atom.
Note that since positronium
consists of a particle and its antiparticle,
$\ab^e_T$ effects cancel completely
providing a clean bound on CPT even effects.
Muonium interferometry \cite{kirch} could offer the analogous sensitivity
for muon coefficients.

The development of cold antihydrogen
provides unique opportunities
to perform tests with neutral antimatter
including some unique tests of Lorentz and CPT symmetry.
A proposal already exists for tests of Lorentz symmetry
based on antihydrogen spectroscopy \cite{eexpt2,antihspec}.
Various ideas also exist for measuring the gravitational acceleration
of antihydrogen
including tests involving
atom traps \cite{antitrap},
atom interferometry \cite{antiinterf,antihinterf},
and free fall from ion traps \cite{antiion}.
A suggestions has also been made
to perform such tests in space,
which would provide further improved sensitivity \cite{antiion}.
Such experiments have the potential
to provide sensitivity to combinations of electron and proton coefficients.
 



\end{document}